\documentclass{article}

\usepackage[final,nonatbib]{neurips_2022}
\pdfoutput=1

\usepackage{mathtools}
\DeclarePairedDelimiter\ceil{\lceil}{\rceil}
\usepackage[utf8]{inputenc} 
\usepackage[T1]{fontenc}    
\usepackage{hyperref}       
\usepackage{url}            
\usepackage{booktabs}       
\usepackage{amsfonts}       
\usepackage{nicefrac}       
\usepackage{microtype}      
\usepackage{xcolor}         
\usepackage[utf8]{inputenc} 
\usepackage[T1]{fontenc}    
\usepackage{hyperref}       
\usepackage{url}            
\usepackage{booktabs}       
\usepackage{amsfonts}       
\usepackage{nicefrac}       
\usepackage{microtype}      
\usepackage{xcolor}         
\usepackage{comment}
\usepackage{graphicx}
\usepackage{caption}
\usepackage{subcaption}
\usepackage{algorithm}
\usepackage{algpseudocode}

\usepackage{amsmath,amsfonts,amssymb}
\usepackage{amsthm}
\usepackage{multirow}
\usepackage{graphicx}
\usepackage{cleveref}
\usepackage{amsthm}
\usepackage{makecell}
\usepackage[english]{babel}
\usepackage{blindtext}
\usepackage{xcolor}
\usepackage{subcaption}
\usepackage{graphicx}
\theoremstyle{definition}
\newtheorem{definition}{Definition}[section]

\newtheorem{theorem}{Theorem}

\theoremstyle{remark}

\title{Differentially Private CutMix for \\Split Learning with Vision Transformer
}

\author{%
  Seungeun Oh, Sihun Baek, Hyelin Nam, Seong-Lyun Kim\thanks{Corresponding author.} \\
  Yonsei University\\
  \texttt{\{seoh,shbaek,hlnam,slkim\}}\\
  \texttt{@ramo.yonsei.ac.kr} \\
  \And
  Jihong Park \\
  Deakin University\\
  \texttt{jihong.park}\\
  \texttt{@deakin.edu.au}\\
  \And
  Praneeth Vepakomma, Ramesh Raskar  \\
  Massachusetts Institute of Technology\\
  \texttt{\{vepakom,raskar\}}\\
  \texttt{@mit.edu}\\
  \And
  Mehdi Bennis  \\
  University of Oulu\\
  \texttt{mehdi.bennis}\\
  \texttt{@oulu.fi}\\ 
}

\begin{document}

\maketitle

\begin{abstract}
Recently, vision transformer (ViT) has started to outpace the conventional CNN in computer vision tasks. Considering privacy-preserving distributed learning with ViT, federated learning (FL) communicates models, which becomes ill-suited due to ViT's large model size and computing costs. Split learning (SL) detours this by communicating smashed data at a cut-layer, yet suffers from data privacy leakage and large communication costs caused by high similarity between ViT's smashed data and input data. Motivated by this problem, we propose \textit{DP-CutMixSL}, a differentially private (DP) SL framework by developing \textit{DP patch-level randomized CutMix (DP-CutMix)}, a novel privacy-preserving inter-client interpolation scheme that replaces randomly selected patches in smashed data. By experiment, we show that DP-CutMixSL not only boosts privacy guarantees and communication efficiency, but also achieves higher accuracy than its Vanilla SL counterpart. Theoretically, we analyze that DP-CutMix amplifies R\'enyi DP (RDP), which is upper-bounded by its Vanilla Mixup counterpart.
\end{abstract}

\section{Introduction}

\paragraph{Motivation: Privacy-Preserving Distributed ML for ViT} 
Edge devices such as phones, cameras, and e-health wearables generate the sheer amount of fresh data~\cite{park2019wireless}. To exploit these user data for machine learning (ML) without violating data privacy, federated learning (FL) is gaining increasing attention, which keeps raw data locally stored while only exchanging and averaging model parameters across devices~\cite{li2020federated,kairouz2021advances}. In particular, FL has been notably successful in computer vision tasks with the \emph{de facto} standard convolutional neural network (CNN) architectures~\cite{liu2020fedvision,he2021fedcv}. However, recently vision transformer (ViT) has been aggressively taking over the throne of CNN~\cite{vaswani2017attention}, questioning the effectiveness of FL. In fact, ViT is often larger than CNN, and this bodes ill for FL by imposing excessive energy and communication burdens on devices~\cite{konevcny2016federated, singh2019detailed}. Alternatively, split learning (SL) can cope with large models via model partitioning~\cite{gupta2018distributed,vepakomma2018split}. In SL, each device locally stores only a tiny fraction of the entire model, and offloads the rest to a parameter server, between which devices exchange their cut-layer forward activations with the server, referred to as \emph{smashed data}. Notwithstanding, ViT commonly lacks pooling and convolutional layers~\cite{vaswani2017attention, devlin2018bert, brown2020language}, making smashed data similar to their raw data as visualized in Fig. 4 of Appendix A. This may entail huge costs and privacy leakage as opposed to its counterpart SL with CNN~\cite{DBLP:journals/corr/abs-2004-12088,thapa2020splitfed,gao2021evaluation}.


\paragraph{Contributions: DP-CutMixSL} To address the aforementioned issues, inspired from the patchfied smashed data in ViT~\cite{vaswani2017attention} and the CutMix technique~\cite{yun2019cutmix}, we propose \emph{DP-CutMixSL} \footnotemark, a differentially private (DP) SL framework with ViT via patch-level randomized CutMix. As Fig. 1 demonstrates, following the Gaussian DP mechanism~\cite{abadi2016deep,gil2013renyi,mironov2017renyi}, each device in DP-CutMixSL first injects random Gaussian noise into smashed data, followed by punching randomly selected patches, yielding \emph{Cutout smashed data} as analogous to those of Cutout~\cite{devries2017improved}. These Cutout smashed data are uploaded to and put together by the server, resulting in \emph{DP-CutMix smashed data} that continue feed-forward propagation. Compared to SL with the Gaussian DP mechanism (\emph{DP-SL}), we theoretically prove that the proposed randomized CutMix operation in DP-CutMixSL amplifies the DP guarantee of smashed data, by up to its upper-bound baseline \emph{DP-MixSL} obtained by replacing CutMix with Mixup~\cite{zhang2017mixup} that simply superimposes the entire patches from each of different smashed data. By experiment, we show that DP-CutMixSL achieves the highest accuracy, followed by DP-MixSL and DP-SL. It is worth noting that while most of the existing works apply Cutout and CutMix at pixel levels for intra-dataset interpolations~\cite{zhao2021intra,zhang2021intra}, we utilize them at patch levels for privacy-preserving inter-dataset interpolations across different devices, i.e., privacy-preserving distributed ML.

\footnotetext{An early version of this work was presented at FL-IJCAI 2022~\cite{baek2022visual}. Compared to~\cite{baek2022visual} proposing CutMixSL and focusing its communication efficiency, this work proposes DP-CutMixSL while studying its DP analysis and the privacy-accuracy trade-off.}


\section{DP-CutMixSL: Patch-Level Randomized CutMix Operations for ViT}


\begin{figure*}[t]
\centering
\includegraphics[width=\textwidth]{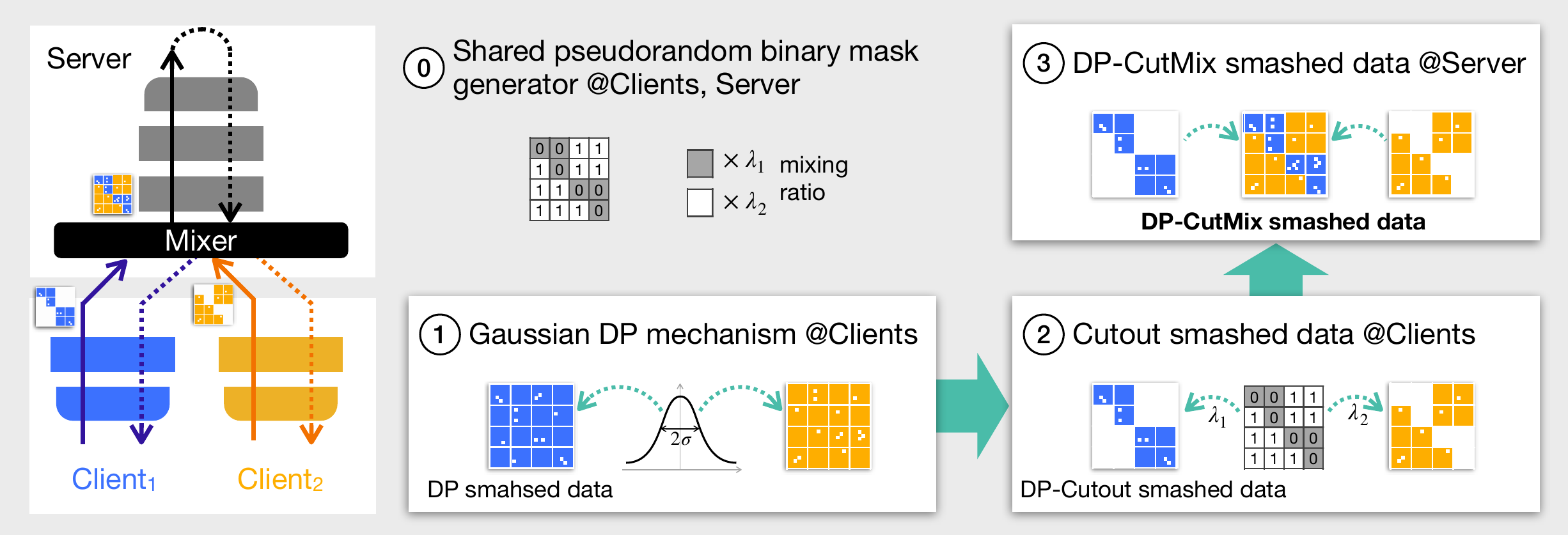}
\caption{A schematic illustration of DP-CutMixSL operations.}
\end{figure*}

The major difference between ViT and CNN can be summarized as follows: 
i) As shown in Fig.~4 of Appendix A, ViT has less feature distortion for the input data of the hidden representation (i.e. smashed data) due to the absence of a pooling layer, ii) Due to its own self-attention mechanism driven by embedding process, ViT captures global spatial information whereas CNN focuses on local spatial information, iii) The above operations of ViT run at patch-level.

At first, i) implies that regularization of the hidden representation in ViT is as efficient as in the input data. Conversely, the mutual information about the input data of the hidden representation is high, leading to data privacy leakage. Next, due to the property of ViT to learn global spatial information mentioned in ii), ViT has more robustness to large-scale noise applied to the fraction of the image~\cite{naseer2021intriguing}, thereby it is suitable for Cutout~\cite{devries2017improved} or CutMix regularization. Finally, iii) suggests the possibility of a patch-scale regularizer. Integrating the above yields a common solution, patch-level randomized CutMix of hidden representations, short for \textit{patch CutMix}. 

Let $i$ and $\mathbf{C}$ be a subscript for a client and a set of clients, respectively. 
As observed in Fig.~1, a mixer, which may be a third-party entity, first generates random sequences $M_i$ with the mixing ratio $\lambda_i \in [0,1]$ $\forall i\in \mathbf{C}$ following a Dirichlet multinomial distribution~\cite{bishop2007discrete}, where $\sum_{i} \lambda_i = 1$. For instance, if a smashed data $s_i$ consists of $N$ patches, $M_i$ is a random binary sequence of length $N$ to control the on-off of each patch, and its non-zero element is $\ceil{\lambda_i\cdot N}$.

Then, the $i$-th client acquires the Cutout smashed data $\bar{s}_i$ by masking the smashed data ($\bar{s}_i=M_i\odot s_i$), obtained by passing the input data through the lower model segment, via the random sequence downloaded from the mixer. When the Cutout smashed data and its label are uploaded to the server, we assume that a gaussian mechanism is applied to them, generating the following DP-Cutout smashed data and label containing white gaussian noise of $N_s$ and $N_y$, respectively:
\begin{align}
    \bar{s}'_{i} &= \bar{s}_{i}+N_s = M_i\odot s_i + N_s, \\ 
    \bar{y}'_{i} &= \bar{y}_{i}+N_y.
\end{align}

The server aggregates DP-Cutout smashed data from all clients and generates \textit{DP-CutMix smashed data} in the following way:
\begin{equation}
    \tilde{s}'_{i,j} = \bar{s}'_i + \bar{s}'_j,\quad\tilde{y}'_{i,j} = \lambda_i\cdot \bar{y}'_i + \lambda_j\cdot \bar{y}'_j,\quad\text{for}\;\, j\neq i.
\end{equation}


Next, the rest of DP-CutMixSL's operation, equal to that of Vanilla SL, performing FP \& BP on the server-side model follows. The said operation of DP-CutMixSL is detailed by the pseudo code of Algorithm 1. 
Fig.~2 also provides image samples of smashed data as well as input data to which the proposed patch CutMix is applied compared to those of Mixup and Vanilla CutMix.

\begin{algorithm}[t]
\caption{DP-CutMixSL} 
\label{alg:appen}
\begin{algorithmic}
\State \textbf{requirements:} $w=[w_{c,i},w_s]^T$ ($w_{c,i}$: lower model segment, $w_s$: upper model segment)

\hspace{42pt} $\eta$: learning rate
\While{$w$ not converged}
\State {\textbf{/*Runs on mixer*/}}               
    \State {samples \{$a_1,.., a_n$\} $\sim$ Dir($\bar{\alpha}$)}
    \State {generates pseudo random sequences $M_i$ for all $i$}  \emph{\hfill$\triangleright$ Pseudorandom binary mask generation}
    \State {unicasts $M_i$ to $i$-th client for all $i$}
    \State
\State {\textbf{/*Runs on client $i\in\mathbf{C}$*/}}
    \State {generates smashed data $s_i$ by passing input data $x_i$ through $w_{c,i}$}
    \State {produces $\bar{s}_i$ by masking $s_i$ via $M_i$}  \emph{\hfill$\triangleright$ Cutout smashed data}
    \State {produces $\bar{s}'_i$ by applying Gaussian mechanism}
    \emph{\hfill$\triangleright$ DP-Cutout smashed data}
    \State {uploads $\bar{s}'_i$ to the server}
    \State
\State {\textbf{/*Runs on server*/}}
    \State {produces $\tilde{s}'_i$ via $\bar{s}'_i$ aggregation for all $i$}
    \emph{\hfill$\triangleright$ DP-CutMix smashed data}    
    \State {generates loss $\sum_i{L_i}$ by passing $\tilde{s}'_i$ through $w_s$ in parallel}
    \State {updates $w_s$ via $w_s\leftarrow w_s-\eta\cdot\nabla_{w_s}{(\sum_i{L_i}})$}
    \emph{\hfill$\triangleright$ Upper model segment update}
    \State {unicasts $i$-th cut-layer gradient to $i$-th client for all $i$}
\State
\State {\textbf{/*Runs on client $i\in\mathbf{C}$*/}}    
    \State {updates $w_{c,i}$ via $w_{c,i}\leftarrow  w_{c,i}-\eta\cdot\nabla_{w_{c,i}}{(\sum_i{L_i}})$}  \emph{\hfill$\triangleright$ Lower model segment update}
    \EndWhile
    \end{algorithmic}
\end{algorithm}

\begin{figure*}[t]
   \begin{subfigure}[h]{0.49\textwidth}
    \includegraphics[clip,width=0.9\textwidth]{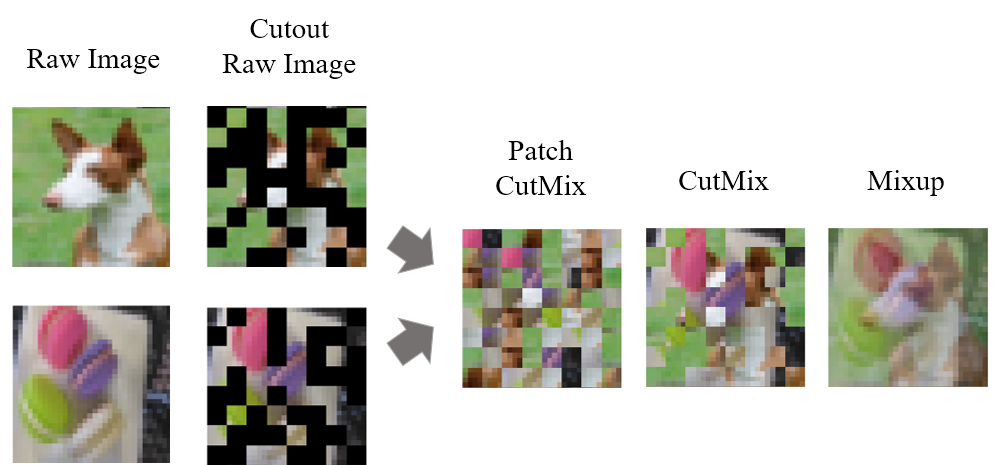}
    \centering
    \caption{Raw images.}
    \label{fig:examp1}
    \end{subfigure}
   \begin{subfigure}[h]{0.49\textwidth}
    \includegraphics[clip,width=0.9\textwidth]{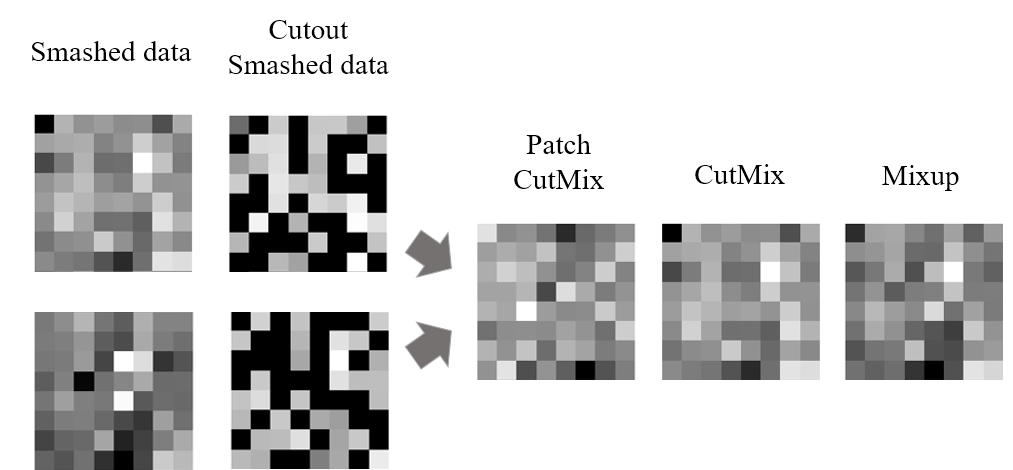}
    \centering
    \caption{Smashed data.}
    \label{fig:examp2}
    \end{subfigure}
    \caption{Examples of data obtained by performing various interpolation schemes on (a) raw image and (b) smashed data.}
\end{figure*}

As a result, DP-CutMixSL can benefit both in terms of privacy leakage and communication cost, in a way that only fraction of the smashed data is shared to the server, even ejected with gaussian noise. Note that random sequences used for smashed data masking are mutually exclusive and collectively exhaustive at the patch-level, so that there are no blank patches in DP-CutMix smashed data. 

\section{DP-CutMixSL: Differential Privacy Analysis}


Let $\mathcal{D} = \{(s_1,y_1), .. , (s_n,y_n)\}$ be a set consisting of $n$ clients' pairs of smashed data $s_i \in \mathbb{R}^{N\times P^2 \times C}$ and the corresponding label $y \in \mathbb{R}^L$ is a one-hot encoding vector, where $P$ denotes the size of patch, respectively. 
We assume that the each element of the smashed data and ground-truth label is upper bounded as follows: $s_i \in [0,\Delta]^{D_s}$ and $y_i \in [0,1]^{D_y}$, where $D_s=NP^2C$ and $D_y=L$. In addition, $\lambda_i$ is the mixing ratio of the $i$-th client, and $N_s$ and $N_y$ are white gaussian noise with dimensions $D_s$ and $D_y$, respectively, i.e., $N_s \sim N(0,\sigma^2_sI_{D_s})$ and $N_y \sim N(0, \sigma^2_yI_{D_y})$ for some $(\sigma_s,\sigma_y)$. Then, we derive the \emph{R\'enyi differential privacy (RDP)}~\cite{mironov2017renyi} of the proposed DP-CutMixSL, which is compared with those of DP-SL and DP-MixSL as follows.
\begin{theorem}
For a given order $\alpha$, the RDP privacy budgets $\epsilon_o(\alpha)$, $\epsilon_{Mix}(\alpha)$, and $\epsilon_{CutMix}(\alpha)$ of DP-SL, DP-MixSL and DP-CutMixSL satisfy $\epsilon_{Mix}(\alpha) \leq \epsilon_{CutMix}(\alpha) \leq \epsilon_{o}(\alpha)$ where:
\begin{align}
    \epsilon_o(\alpha) &= \frac{\alpha}{2}{\left(\frac{\Delta^2 D_s}{\sigma^2_s}+\frac{D_y}{\sigma^2_y}\right)},\\
    \epsilon_{Mix}(\alpha) &= \frac{\alpha\left(\max_{i\in\mathbf{C}}\lambda_i\right)^2}{2}{\left(\frac{\Delta^2 D_s}{\sigma^2_s}+\frac{D_y}{\sigma^2_y}\right)},\\
        \epsilon_{CutMix}(\alpha) &= \frac{\alpha\left(\max_{i\in\mathbf{C}}\lambda_i\right)}{2}{\left(\frac{\Delta^2 D_s}{\sigma^2_s}+\frac{D_y\left(\max_{i\in\mathbf{C}}\lambda_i\right)}{\sigma^2_y}\right)}.
\end{align}
\emph{Sketch of Proof.}
For each technique, we derive its output representation, followed by calculating the RDP bound using the output via the R\'enyi divergence formula for a multivariate Gaussian distribution~\cite{gil2013renyi}. Applying this to both the smashed data and the label and combining them via the sequential composition rule completes the proof. The details are deferred to Appendix B.  \hfill $\blacksquare$
\end{theorem}
\vspace{-3pt}
Since $\lambda_i \in [0,1]$, DP-MixSL achieves the highest RDP guarantee (i.e., tightest RDP bound) compared to DP-CutMixSL and DP-SL, with the help of the inherent distortion property of interpolations~\cite{koda2021airmixml, borgnia2021dp, lee2019synthesizing}. It is worth noting that the case only when $\max_{i\in\mathbf{C}}\lambda_i= 1$, i.e., a single client scenario with $|\mathbf{C}|=1$, the equality conditions $\epsilon_{Mix}(\alpha) = \epsilon_{CutMix}(\alpha) = \epsilon_{o}(\alpha)$ hold. In other words, none of equality does not hold for multi clients. Note here that we focus only on the RDP guarantees of smashed data. Smashed data are vulnerable to reconstruction attacks~\cite{gawron2022feature}, threatening the privacy of raw data, which is discussed in Appendix C. In addition, while we focus on the label privacy in the FP, the label privacy can also be leaked from gradients in the BP via white-box attacks. To prevent this, BP label privacy guaranteeing method such as GradPerturb~\cite{yang2022differentially} can additionally be integrated, which is deferred to future research.

\vspace{-5pt}
\section{Numerical Evaluation}
\vspace{-5pt}
In this section, we measure the accuracy, RDP bound ($\epsilon$), and scalability of DP-CutMixSL compared to those of SplitFed~\cite{thapa2020splitfed}, DP-SL, DP-MixSL, and etc.
In Table 1, both the CIFAR-10 dataset~\cite{Krizhevsky09learningmultiple} and the Fashion-MNIST dataset~\cite{xiao2017fashion} are utilized under three types of models: ViT-tiny~\cite{touvron2021training}, PiT-tiny~\cite{DBLP:journals/corr/abs-2103-16302}, and VGG-16~\cite{simonyan2014very}. Here, PiT is a transformer structure equipped with a pooling layer and is a model between ViT and CNN. For all SL algorithms, we assume that the cut-layer is located after embedding process. Other parameters especially for RDP calculation are as follows: patch size $N=64$, $D_s=20$, $D_y=10$, $\Delta=0.2$, $\lambda_i=1/n$ $\forall i$ (uniform), and RDP parameter $\alpha=2$.

\vspace{-3pt}

Table 1 shows the top-1 accuracy for several SL methods including the proposed CutMixSL in a noiseless environment.
As seen at Table 1, except for one case, where SL w. Mixup is used with CIFAR-10 dataset and VGG-16 model, the CutMixSL outperforms other state-of-the-art SL algorithms in terms of top-1 accuracy. This is rooted in the difference between ViT and CNN mentioned in Sec.~2. When learning spatial information, ViT focuses on globality due to the self-attention mechanism, whereas CNN focuses on locality. Hence, a patch CutMix in which certain patches are replaced by patches of other smashed data may cause significant information loss in CNN. On the other hand, interpolation such as mixup is less likely to yield large information loss, thereby CNN and ViT are suitable for mixup and patch CutMix, respectively. Comparing CutMixSL and SL w. Vanilla CutMix in Table 1 proves that patch-level random punching is more efficient than bounding box-based process. This is because the random punching method is a regularizer well suited to ViT, where all operations operate at patch-level, unlike Vanilla CutMix's bounding box where the size is regardless of patch size. From a dropout~\cite{srivastava2014dropout} perspective, the random punching method of CutMixSL is more similar to dropout than mixup or Vanilla CutMix, leading to accuracy gains.

\begin{table}[t]
\caption{Top-1 accuracy of SL-based techniques w.r.t various model types and datasets.}
\centering
    \resizebox{0.8\columnwidth}{!}{
\begin{tabular}{l|rrr|rrr}
\toprule
\multirow{2}{*}{Method ($|\mathbf{C}=10|$)} &  \multicolumn{3}{c|}{Models w/ CIFAR-10} &  \multicolumn{3}{c}{Models w/ Fashion-MNIST}  \\ \cmidrule(l){2-4} \cmidrule(l){5-7}
 & ViT-Tiny & PiT-Tiny & VGG-16 & ViT-Tiny & PiT-Tiny & VGG-16  \\ \midrule
Standalone & 48.84 & 47.77 & 54.97 & 77.65 & 78.21 & 80.12\\
SL~\cite{vepakomma2018split}  & 57.21 & 52.28 & 62.62 &85.68 & 82.35 & 84.39\\
SplitFed~\cite{thapa2020splitfed} & 67.88 & 55.63& 63.98& 89.17 & 84.27 & 87.34 \\\midrule 
Standalone w. Cutout~\cite{devries2017improved} & 53.86 & 50.28 & 56.65 & 88.46 & 86.48 & 88.17 \\ 
SL w. Mixup & 69.23 & 64.89  & \textbf{68.20} & 88.21 & 87.62 & 88.53 \\ 
SL w. Vanilla CutMix  & 71.78 & 58.21 & 33.50 & 87.86 & 86.31& 89.01 \\
CutMixSL (proposed) & \textbf{73.77} &  \textbf{71.26} & 67.53 & \textbf{89.75} & \textbf{89.25} &\textbf{89.45}\\\midrule
\end{tabular}}
\label{table:models}
\end{table}

\begin{figure}[t]
\centering
\begin{subfigure}[h]{0.33\textwidth}
\centering
\includegraphics[width=1\textwidth]{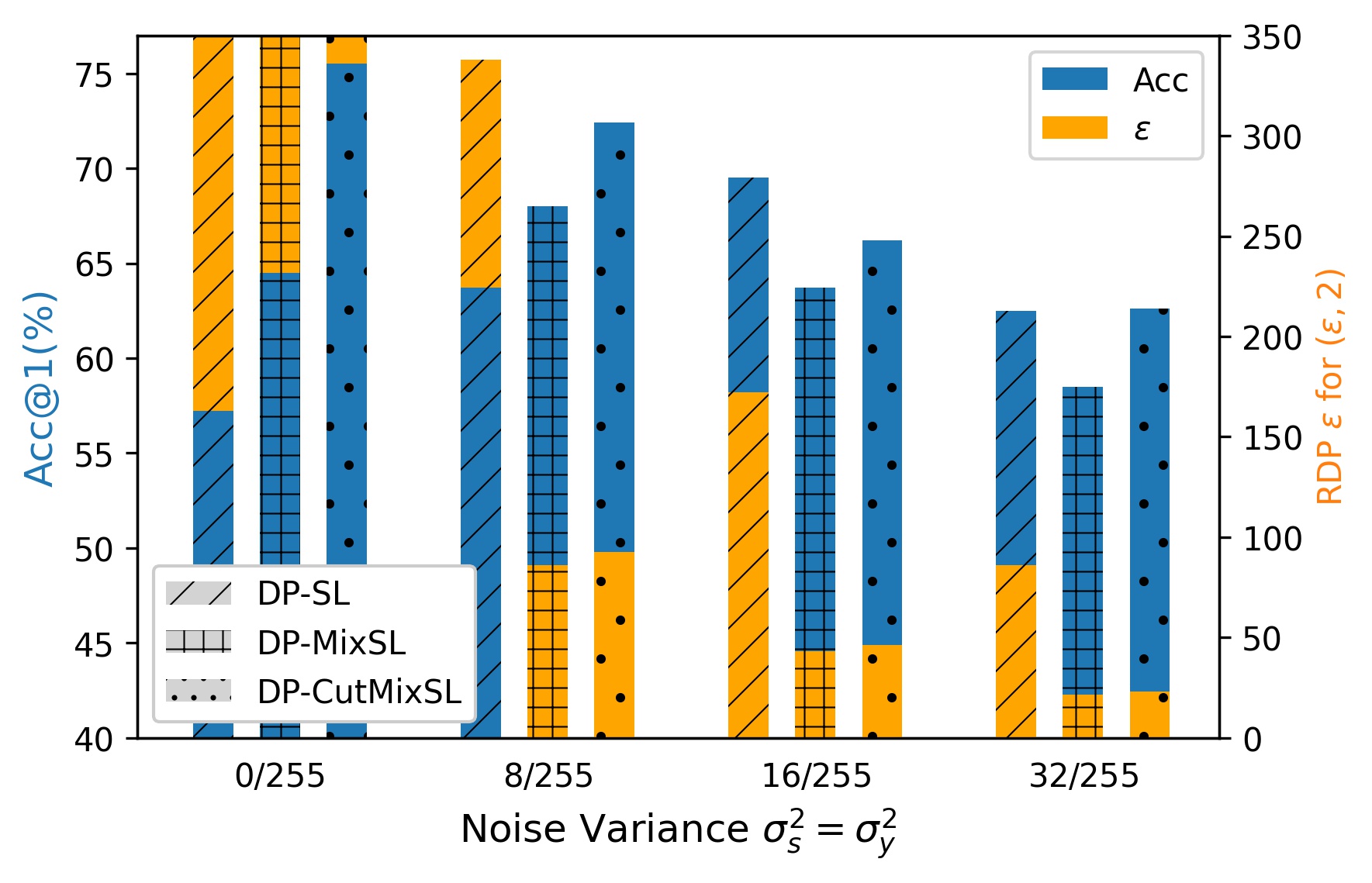}
\caption{Acc-$\epsilon$ per noise variance.}
\label{fig:group size}
\end{subfigure}
\hfill
\begin{subfigure}[h]{0.34\textwidth}
\centering
\includegraphics[width=1\textwidth]{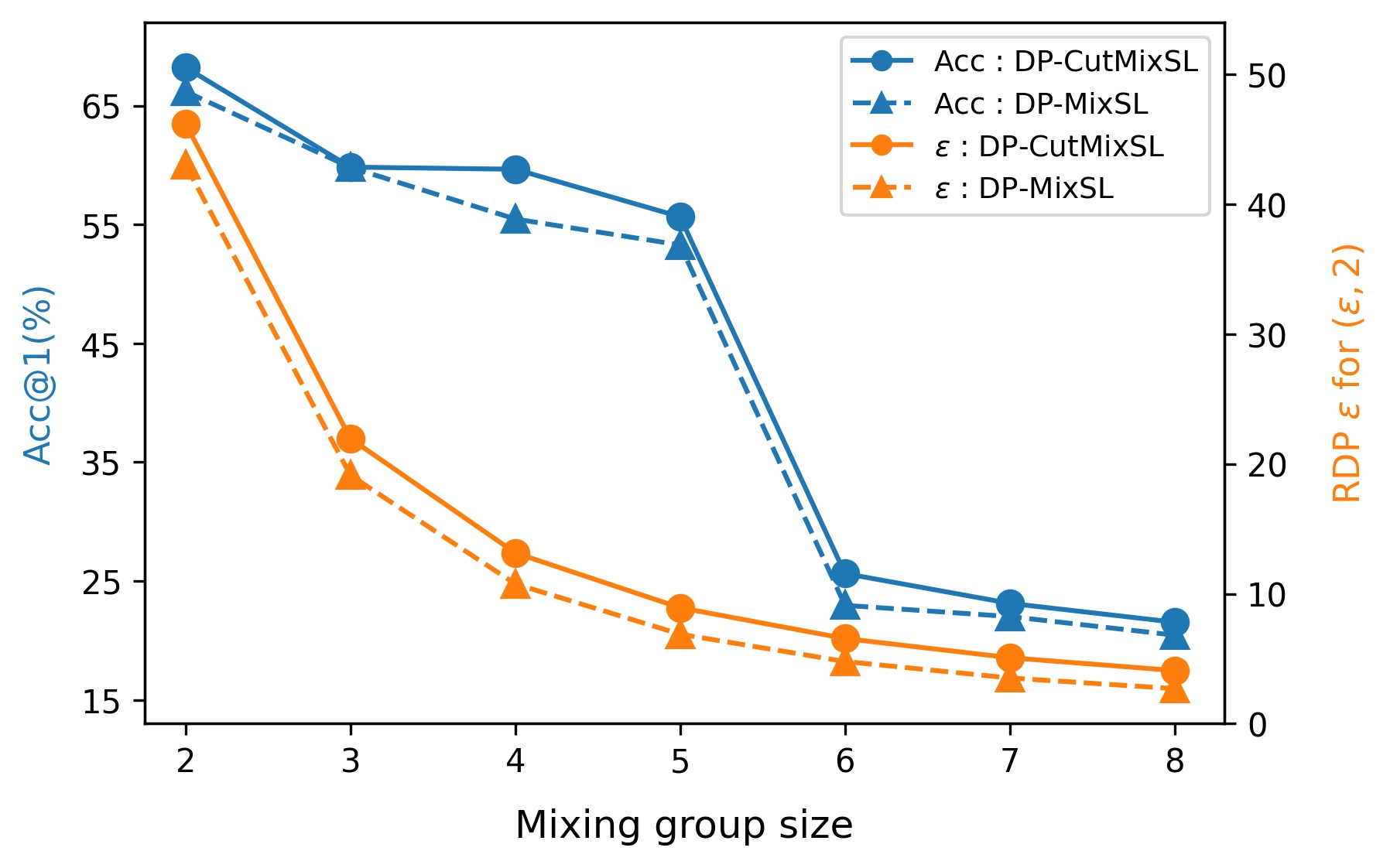}
\caption{Acc-$\epsilon$ of mixing methods.}
\label{fig:noise var}
\end{subfigure}
\begin{subfigure}[h]{0.3\textwidth}
\centering
\includegraphics[width=1\textwidth]{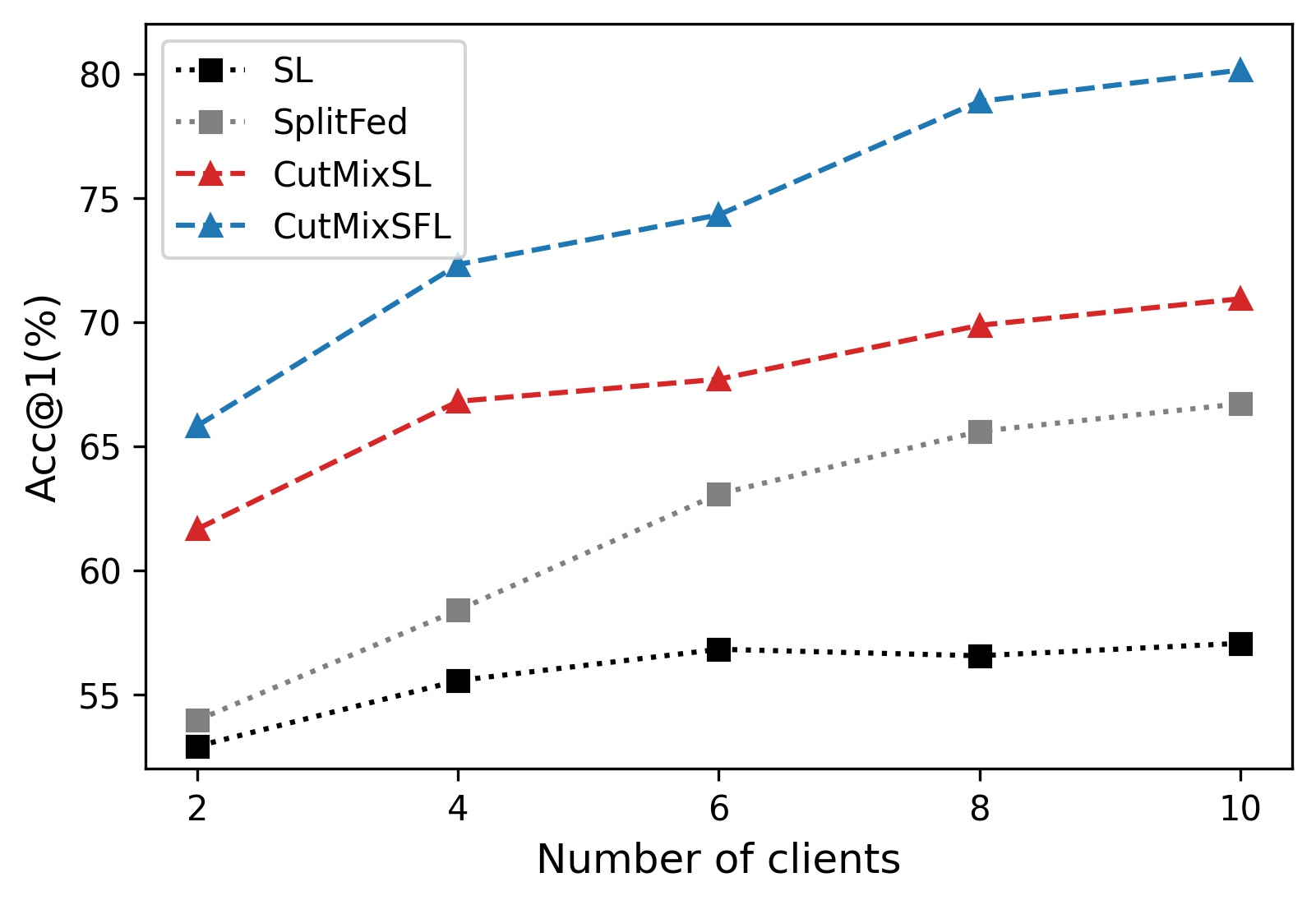}
\caption{Scalability.}
\label{fig:dp comb}
\end{subfigure}
\caption{Accuracy and $\epsilon$ under the CIFAR-10 dataset: (a) accuracy and $\epsilon$ of DP-SL, DP-MixSL, and DP-CutMixSL w.r.t noise variance; (b) accuracy and $\epsilon$ of DP-MixSL and DP-CutMixSL w.r.t the mixing group size; (c) accuracy of various SL-based techniques 
according to number of clients.}
\label{DP}
\end{figure}

Fig. 3a shows the effect of noise variance on accuracy and $\epsilon$. In terms of accuracy, DP-CutMixSL is the best, except when the noise variance is $16/255$. Compared to DP-MixSL, DP-CutMixSL has superior performance in all cases. Looking at $\epsilon$, however, DP-CutMixSL has a tighter RDP bound compared to DP-SL, but has a larger $\epsilon$ in comparison with DP-MixSL, showing the \textit{accuracy-privacy trade-off}.
In Fig. 3b, the size of a mixing group, a set of clients taking a mixup or CutMix, varies both in DP-CutMixSL and DP-MixSL. In both DP-CutMixSL and DP-MixSL, the accuracy and $\epsilon$ decrease as the mixing group size increases, also resulting in the accuracy-privacy trade-off. The decrease in $\epsilon$ according to the mixing group size can be explained by the "Hiding in the crowd" effect~\cite{jeong2020hiding}, and the decrease in accuracy is also explained in connection with the information loss mentioned above. Fig. 3c shows the curve of scalability, in terms of accuracy increase according to the number of clients. In Fig. 3c, all SL techniques including CutMixSL guarantee scalability when the client increases from 2 to 10, and the accuracy of CutMixSFL, which introduced SplitFed's weight averaging of lower model segment to CutMixSL, is further improved.
\vspace{-5pt}

\section{Concluding Remarks}
\vspace{-5pt}
In this work, we proposed DP-CutMixSL for privacy-preserving distributed ML for ViT by exploiting CutMix for inter-dataset interpolations. We theoretically analyzed its DP guarantee, and numerically showed its achieving the highest accuracy compared to two baselines, DP-SL and DP-MixSL. While we focus only on generating a single CutMix output for two or more inputs, it is possible to generate multiple outputs by changing the mixing ratio for interpolations with finer resolution as done in~\cite{oh2022locfedmix}, which could be an interesting topic for future study. Furthermore, based on the simulation results showing the effectiveness of the proposed method even in the presence of several pooling and convolutional layers, it is worth investigating other patchified architectures in different domains for future research. 

\begin{ack}
\vspace{-10pt}
This work was supported by Institute of Information \& communications Technology Planning \& Evaluation (IITP) grant funded by the Korea government (MSIT) (No. 2021-0-02208, Sub-THz Augmented Routing and Transmission for 6G).Evaluation (IITP) grant funded by the Korea government (MSIT) (No. 2021-0-02208, Sub-THz Augmented Routing and Transmission for 6G). 
\end{ack}

\newpage

\begin{thebibliography}{10}

\bibitem{park2019wireless}
Jihong Park, Sumudu Samarakoon, Mehdi Bennis, and M{\'e}rouane Debbah.
\newblock Wireless network intelligence at the edge.
\newblock {\em Proceedings of the IEEE}, 107(11):2204--2239, 2019.

\bibitem{li2020federated}
Tian Li, Anit~Kumar Sahu, Ameet Talwalkar, and Virginia Smith.
\newblock Federated learning: Challenges, methods, and future directions.
\newblock {\em IEEE Signal Processing Magazine}, 37(3):50--60, 2020.

\bibitem{kairouz2021advances}
Peter Kairouz, H~Brendan McMahan, Brendan Avent, Aur{\'e}lien Bellet, Mehdi
  Bennis, Arjun~Nitin Bhagoji, Kallista Bonawitz, Zachary Charles, Graham
  Cormode, Rachel Cummings, et~al.
\newblock Advances and open problems in federated learning.
\newblock {\em Foundations and Trends{\textregistered} in Machine Learning},
  14(1--2):1--210, 2021.

\bibitem{liu2020fedvision}
Yang Liu, Anbu Huang, Yun Luo, He~Huang, Youzhi Liu, Yuanyuan Chen, Lican Feng,
  Tianjian Chen, Han Yu, and Qiang Yang.
\newblock Fedvision: An online visual object detection platform powered by
  federated learning.
\newblock In {\em Proceedings of the AAAI Conference on Artificial
  Intelligence}, volume~34, pages 13172--13179, 2020.

\bibitem{he2021fedcv}
Chaoyang He, Alay~Dilipbhai Shah, Zhenheng Tang, Di~Fan1Adarshan~Naiynar
  Sivashunmugam, Keerti Bhogaraju, Mita Shimpi, Li~Shen, Xiaowen Chu, Mahdi
  Soltanolkotabi, and Salman Avestimehr.
\newblock Fedcv: a federated learning framework for diverse computer vision
  tasks.
\newblock {\em arXiv preprint arXiv:2111.11066}, 2021.

\bibitem{vaswani2017attention}
Ashish Vaswani, Noam Shazeer, Niki Parmar, Jakob Uszkoreit, Llion Jones,
  Aidan~N Gomez, {\L}ukasz Kaiser, and Illia Polosukhin.
\newblock Attention is all you need.
\newblock {\em Advances in neural information processing systems}, 30, 2017.

\bibitem{konevcny2016federated}
Jakub Kone{\v{c}}n{\`y}, H~Brendan McMahan, Felix~X Yu, Peter Richt{\'a}rik,
  Ananda~Theertha Suresh, and Dave Bacon.
\newblock Federated learning: Strategies for improving communication
  efficiency.
\newblock {\em arXiv preprint arXiv:1610.05492}, 2016.

\bibitem{singh2019detailed}
Abhishek Singh, Praneeth Vepakomma, Otkrist Gupta, and Ramesh Raskar.
\newblock Detailed comparison of communication efficiency of split learning and
  federated learning.
\newblock {\em arXiv preprint arXiv:1909.09145}, 2019.

\bibitem{gupta2018distributed}
Otkrist Gupta and Ramesh Raskar.
\newblock Distributed learning of deep neural network over multiple agents.
\newblock {\em Journal of Network and Computer Applications}, 116:1--8, 2018.

\bibitem{vepakomma2018split}
Praneeth Vepakomma, Otkrist Gupta, Tristan Swedish, and Ramesh Raskar.
\newblock Split learning for health: Distributed deep learning without sharing
  raw patient data.
\newblock {\em arXiv preprint arXiv:1812.00564}, 2018.

\bibitem{devlin2018bert}
Jacob Devlin, Ming-Wei Chang, Kenton Lee, and Kristina Toutanova.
\newblock Bert: Pre-training of deep bidirectional transformers for language
  understanding.
\newblock {\em arXiv preprint arXiv:1810.04805}, 2018.

\bibitem{brown2020language}
Tom Brown, Benjamin Mann, Nick Ryder, Melanie Subbiah, Jared~D Kaplan, Prafulla
  Dhariwal, Arvind Neelakantan, Pranav Shyam, Girish Sastry, Amanda Askell,
  et~al.
\newblock Language models are few-shot learners.
\newblock {\em Advances in neural information processing systems},
  33:1877--1901, 2020.

\bibitem{DBLP:journals/corr/abs-2004-12088}
Chandra Thapa, Mahawaga Arachchige~Pathum Chamikara, and Seyit Camtepe.
\newblock Splitfed: When federated learning meets split learning.
\newblock {\em CoRR}, abs/2004.12088, 2020.

\bibitem{thapa2020splitfed}
Chandra Thapa, Mahawaga Arachchige~Pathum Chamikara, Seyit Camtepe, and Lichao
  Sun.
\newblock Splitfed: When federated learning meets split learning.
\newblock {\em arXiv preprint arXiv:2004.12088}, 2020.

\bibitem{gao2021evaluation}
Yansong Gao, Minki Kim, Chandra Thapa, Sharif Abuadbba, Zhi Zhang, Seyit
  Camtepe, Hyoungshick Kim, and Surya Nepal.
\newblock Evaluation and optimization of distributed machine learning
  techniques for internet of things.
\newblock {\em IEEE Transactions on Computers}, 2021.

\bibitem{yun2019cutmix}
Sangdoo Yun, Dongyoon Han, Seong~Joon Oh, Sanghyuk Chun, Junsuk Choe, and
  Youngjoon Yoo.
\newblock Cutmix: Regularization strategy to train strong classifiers with
  localizable features.
\newblock In {\em Proceedings of the IEEE/CVF International Conference on
  Computer Vision}, pages 6023--6032, 2019.

\bibitem{abadi2016deep}
Martin Abadi, Andy Chu, Ian Goodfellow, H~Brendan McMahan, Ilya Mironov, Kunal
  Talwar, and Li~Zhang.
\newblock Deep learning with differential privacy.
\newblock In {\em Proceedings of the 2016 ACM SIGSAC conference on computer and
  communications security}, pages 308--318, 2016.

\bibitem{gil2013renyi}
Manuel Gil, Fady Alajaji, and Tamas Linder.
\newblock R{\'e}nyi divergence measures for commonly used univariate continuous
  distributions.
\newblock {\em Information Sciences}, 249:124--131, 2013.

\bibitem{mironov2017renyi}
Ilya Mironov.
\newblock R{\'e}nyi differential privacy.
\newblock In {\em 2017 IEEE 30th computer security foundations symposium
  (CSF)}, pages 263--275. IEEE, 2017.

\bibitem{devries2017improved}
Terrance DeVries and Graham~W Taylor.
\newblock Improved regularization of convolutional neural networks with cutout.
\newblock {\em arXiv preprint arXiv:1708.04552}, 2017.

\bibitem{zhang2017mixup}
Hongyi Zhang, Moustapha Cisse, Yann~N Dauphin, and David Lopez-Paz.
\newblock mixup: Beyond empirical risk minimization.
\newblock {\em arXiv preprint arXiv:1710.09412}, 2017.

\bibitem{zhao2021intra}
Caidan Zhao and Yang Lei.
\newblock Intra-class cutmix for unbalanced data augmentation.
\newblock In {\em 2021 13th International Conference on Machine Learning and
  Computing}, pages 246--251, 2021.

\bibitem{zhang2021intra}
Lianbo Zhang, Shaoli Huang, and Wei Liu.
\newblock Intra-class part swapping for fine-grained image classification.
\newblock In {\em Proceedings of the IEEE/CVF Winter Conference on Applications
  of Computer Vision}, pages 3209--3218, 2021.

\bibitem{baek2022visual}
Sihun Baek, Jihong Park, Praneeth Vepakomma, Ramesh Raskar, Mehdi Bennis, and
  Seong-Lyun Kim.
\newblock Visual transformer meets cutmix for improved accuracy, communication
  efficiency, and data privacy in split learning.
\newblock {\em arXiv preprint arXiv:2207.00234}, 2022.

\bibitem{naseer2021intriguing}
Muhammad~Muzammal Naseer, Kanchana Ranasinghe, Salman~H Khan, Munawar Hayat,
  Fahad Shahbaz~Khan, and Ming-Hsuan Yang.
\newblock Intriguing properties of vision transformers.
\newblock {\em Advances in Neural Information Processing Systems}, 34, 2021.

\bibitem{bishop2007discrete}
Yvonne~M Bishop, Stephen~E Fienberg, and Paul~W Holland.
\newblock {\em Discrete multivariate analysis: theory and practice}.
\newblock Springer Science \& Business Media, 2007.

\bibitem{koda2021airmixml}
Yusuke Koda, Jihong Park, Mehdi Bennis, Praneeth Vepakomma, and Ramesh Raskar.
\newblock Airmixml: Over-the-air data mixup for inherently privacy-preserving
  edge machine learning.
\newblock {\em arXiv preprint arXiv:2105.00395}, 2021.

\bibitem{borgnia2021dp}
Eitan Borgnia, Jonas Geiping, Valeriia Cherepanova, Liam Fowl, Arjun Gupta,
  Amin Ghiasi, Furong Huang, Micah Goldblum, and Tom Goldstein.
\newblock Dp-instahide: Provably defusing poisoning and backdoor attacks with
  differentially private data augmentations.
\newblock {\em arXiv preprint arXiv:2103.02079}, 2021.

\bibitem{lee2019synthesizing}
Kangwook Lee, Hoon Kim, Kyungmin Lee, Changho Suh, and Kannan Ramchandran.
\newblock Synthesizing differentially private datasets using random mixing.
\newblock In {\em 2019 IEEE International Symposium on Information Theory
  (ISIT)}, pages 542--546. IEEE, 2019.

\bibitem{gawron2022feature}
Grzegorz Gawron and Philip Stubbings.
\newblock Feature space hijacking attacks against differentially private split
  learning.
\newblock {\em arXiv preprint arXiv:2201.04018}, 2022.

\bibitem{yang2022differentially}
Xin Yang, Jiankai Sun, Yuanshun Yao, Junyuan Xie, and Chong Wang.
\newblock Differentially private label protection in split learning.
\newblock {\em arXiv preprint arXiv:2203.02073}, 2022.

\bibitem{Krizhevsky09learningmultiple}
Alex Krizhevsky.
\newblock Learning multiple layers of features from tiny images.
\newblock Technical report, 2009.

\bibitem{xiao2017fashion}
Han Xiao, Kashif Rasul, and Roland Vollgraf.
\newblock Fashion-mnist: a novel image dataset for benchmarking machine
  learning algorithms.
\newblock {\em arXiv preprint arXiv:1708.07747}, 2017.

\bibitem{touvron2021training}
Hugo Touvron, Matthieu Cord, Matthijs Douze, Francisco Massa, Alexandre
  Sablayrolles, and Herv{\'e} J{\'e}gou.
\newblock Training data-efficient image transformers \& distillation through
  attention.
\newblock In {\em International Conference on Machine Learning}, pages
  10347--10357. PMLR, 2021.

\bibitem{DBLP:journals/corr/abs-2103-16302}
Byeongho Heo, Sangdoo Yun, Dongyoon Han, Sanghyuk Chun, Junsuk Choe, and
  Seong~Joon Oh.
\newblock Rethinking spatial dimensions of vision transformers.
\newblock {\em CoRR}, abs/2103.16302, 2021.

\bibitem{simonyan2014very}
Karen Simonyan and Andrew Zisserman.
\newblock Very deep convolutional networks for large-scale image recognition.
\newblock {\em arXiv preprint arXiv:1409.1556}, 2014.

\bibitem{srivastava2014dropout}
Nitish Srivastava, Geoffrey Hinton, Alex Krizhevsky, Ilya Sutskever, and Ruslan
  Salakhutdinov.
\newblock Dropout: a simple way to prevent neural networks from overfitting.
\newblock {\em The journal of machine learning research}, 15(1):1929--1958,
  2014.

\bibitem{jeong2020hiding}
Eunjeong Jeong, Seungeun Oh, Jihong Park, Hyesung Kim, Mehdi Bennis, and
  Seong-Lyun Kim.
\newblock Hiding in the crowd: Federated data augmentation for on-device
  learning.
\newblock {\em IEEE Intelligent Systems}, 36(5):80--87, 2020.

\bibitem{oh2022locfedmix}
Seungeun Oh, Jihong Park, Praneeth Vepakomma, Sihun Baek, Ramesh Raskar, Mehdi
  Bennis, and Seong-Lyun Kim.
\newblock Locfedmix-sl: Localize, federate, and mix for improved scalability,
  convergence, and latency in split learning.
\newblock In {\em Proceedings of the ACM Web Conference 2022}, pages
  3347--3357, 2022.

\end{thebibliography}
\bibliographystyle{plainnat}

\newpage

\appendix

\section{Image Visualization in CNN and ViT}

\begin{figure*}[h]
\centering
\includegraphics[width=0.85\textwidth]{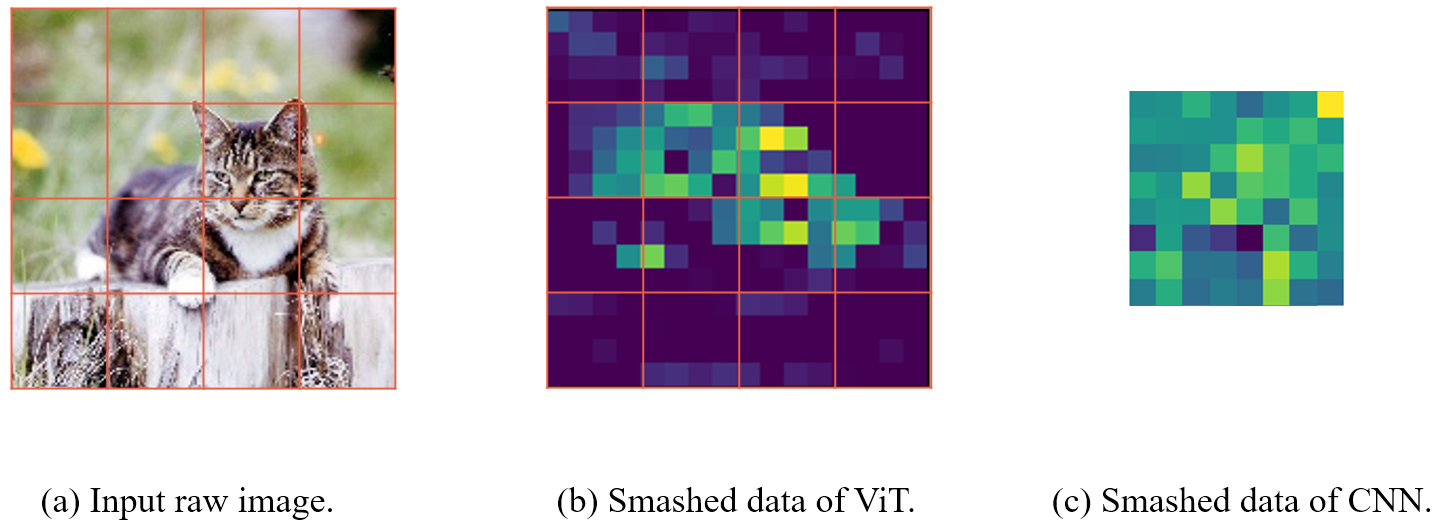}
\caption{Comparison between raw image and smashed data of ViT and CNN.}
\end{figure*}


\section{RDP Analysis}
Before going further, the definition of RDP is as follows:
\begin{definition}[$(\alpha, \epsilon)$-RDP~\cite{mironov2017renyi}]
A randomized mechanism $f : \mathcal{D} \rightarrow \mathcal{R}$ is said to have $\epsilon$-R$\mathrm{\acute{e}}$nyi differential privacy of order $\alpha$, or ($\alpha, \epsilon$)-RDP for short, if for any adjacent $D, D' \in \mathcal{D}$ it holds that 
\begin{align}
D_{\alpha}(f(D)||f(D')) \leq \epsilon
\end{align}
\end{definition}



A strong privacy guarantee implies that one cannot distinguish whether $D$ or $D'$ was used to produce an outcome of mechanism. 

\subsection{DP-SL}
Here, our baseline DP mechanism is Gaussian mechanism. Then, the output of DP-SL, which are the smashed data and its label to which gaussian noise is applied, respectively, is as follows:
\begin{align}
s'_i=s_i+N_s,\label{eq:8}\\
y'_i=y_i+ N_y,
\end{align}
where $N_s \sim N(0,\sigma^2_sI_{D_s})$ and $N_y \sim N(0, \sigma^2_yI_{D_y})$ for some $(\sigma_s,\sigma_y)$. 

By using the Definition B.1 and R\'enyi divergence formula from \cite{gil2013renyi}, the RDP bound of gaussian mechanism $\mathcal{M}$ for DP-SL, $\epsilon_{o}(\alpha)$ can be expressed as:
\begin{align}
\epsilon_{o}(\alpha)=\sup_{\mathcal{D},\mathcal{D'}}D_\alpha(\mathcal{M(D)}||\mathcal{M(D')})=\sup_{\mathcal{D},\mathcal{D'}}\frac{\alpha}{2\sigma^2}{||\mu^\mathcal{D}_X-\mu^{\mathcal{D'}}_X||}^2, 
\label{gaussian}
\end{align}

where $\mathcal{M(D)} \sim N(\mu^\mathcal{D}_X, \sigma_X^2)$ and $\mathcal{M(D')} \sim N(\mu^{\mathcal{D'}}_X, \sigma_X^2)$.

Let $s^{\mathcal{D}}_{(i,k)}$ denote the $k$-th element of smashed data $s^{\mathcal{D}}_i$ in dataset $\mathcal{D}$, where $s^{\mathcal{D}}_i=[s^{\mathcal{D}}_{(i,1)},...,s^{\mathcal{D}}_{(i,D_s)}]$. Then, ${s}'^{\mathcal{D}}_{i}$ obtained by passing $s^{\mathcal{D}}_i$ through (\ref{eq:8}) follows a gaussian distribution with mean $s^{\mathcal{D}}_i$ and variance $\sigma_s^2$ (In element perspective,
${s}'^{\mathcal{D}}_{(i,k)}\sim N(s^{\mathcal{D}}_{(i,k)},\sigma_s^2)$ for $k \in [D_s]$).


For two sets of smashed data $\mathcal{D}$ and $\mathcal{D'}$ where only the $i'$-th smashed data is different, (\ref{gaussian}) becomes:
\begin{align}
\sup_{\mathcal{D},\mathcal{D'}}\frac{\alpha}{2\sigma_s^2}{\|\mu_X^\mathcal{D} - \mu_X^{\mathcal{D}'}\|}^2=\frac{\alpha}{2\sigma_s^2}\sum^{D_s}_{k=1}{(s_{(i',k)}^\mathcal{D}-s_{(i',k)}^{\mathcal{D}'}
)}^2. 
\label{eq:l2}
\end{align}


Considering the element-wise upper bound of DP-SL's smashed data yields the following formula:
\begin{align}
\sum^{D_s}_{k=1}{(s_{(i',k)}^\mathcal{D}-s_{(i',k)}^{\mathcal{D}'}
)}^2\leq\Delta^2\cdot D_s.
\end{align}

Therefore, gaussian mechanism for smashed data in DP-SL is $(\alpha, \epsilon_{o})$-RDP, where 
\begin{align}
\epsilon_{o}(\alpha) =\alpha\frac{\Delta^2\cdot D_s}{2\sigma_s^2}.
\end{align}

Likewise, RDP bound of groud-truth label in DP-SL can be calculated, yielding $\epsilon_{o}(\alpha)$ as below:
 \begin{align}
 \epsilon_{o}(\alpha) = \frac{\alpha}{2}{\left(\frac{\Delta^2 D_s}{\sigma^2_s}+\frac{D_y}{\sigma^2_y}\right)}.
 \label{SL_epi}
 \end{align}

\subsection{DP-MixSL}

The output of DP-MixSL can be expressed as $\hat{s}=\sum_{i=1}^n \lambda_i s'_i$, where $\{\lambda_1,...,\lambda_n\}$ is a set of mixing ratio of each client's smashed data. Also, DP-MixSL executes mixup operation similar to its corresponding label.

Then, for two adjacent datasets $\mathcal{D}$ and $\mathcal{D}'$, (\ref{gaussian}) becomes:
\begin{align}
\sup_{\mathcal{D},\mathcal{D'}}\frac{\alpha}{2\sigma_s^2}{\|\mu_X^\mathcal{D} - \mu_X^{\mathcal{D}'}\|}^2=\frac{\alpha}{2\sigma_s^2}\sum^{D_s}_{k=1}{(\lambda_{i'}(s_{(i',k)}^\mathcal{D}-s_{(i',k)}^{\mathcal{D}'}
))}^2. 
\label{mixup:eq}
\end{align}


This bound can be maximized when the $i'$-th mixing ratio is the largest value of $\lambda_i$ for all $i\in\mathbf{C}$.
In this case, from \ref{mixup:eq}, we have
\begin{align}
\sum^{D_s}_{k=1}{(\lambda_{i'}(s_{(i',k)}^\mathcal{D}-s_{(i',k)}^{\mathcal{D}'}))}^2 \leq  (\max_{i\in \mathbf{C}} \lambda_i)^2 \Delta^2 D_s.
\label{eq:21}
\end{align}

By applying the same process above to the label, we have 
\begin{align}
\epsilon_{Mix}(\alpha) &= \epsilon_o(\alpha)\cdot \left(\max_{i\in\mathbf{C}}\lambda_i\right)^2 \\
&= \frac{\alpha\left(\max_{i\in\mathbf{C}}\lambda_i\right)^2}{2}{\left(\frac{\Delta^2 D_s}{\sigma^2_s}+\frac{D_y}{\sigma^2_y}\right)}.
\label{mix_epi}
\end{align}


\subsection{DP-CutMixSL}
The output of DP-CutMixSL can be represented by $\tilde{s}=\sum_{i=1}^n M_i\odot s'_i$, whereas its operation on label is the same as in DP-MixSL. For two adjacent datasets with only one smashed data different, DP-CutMixSL only needs to calculate bounds only for the element in which the smashed data is masked, in contrast to DP-MixSL, where the smashed data is melted in the whole element.

That is, assuming that the number of 1 elements included in the $i'$-th mask is $N_{i'}$, 
the following inequality is derived from (\ref{gaussian}):
\begin{align}
\sup_{\mathcal{D},\mathcal{D'}}\frac{\alpha}{2\sigma_s^2}{\|\mu_X^\mathcal{D} - \mu_X^{\mathcal{D}'}\|}^2\leq\frac{\alpha}{2\sigma_s^2}N_{i'}\Delta^2=\frac{\alpha}{2\sigma_s^2}\lambda_{i'}D_s\Delta^2. 
\label{eq:cutmix22}
\end{align}

(\ref{eq:cutmix22}) has an upper bound as shown below when $\lambda_{i'}$ is the maximum among $\lambda_i$ $\forall i$:
\begin{align}
\frac{\alpha}{2\sigma_s^2}\lambda_{i'}D_s\Delta^2 \leq (\max_{i\in\mathbf{C}}\lambda_i)\Delta^2 D_s.
\label{eq:cutmix23}
\end{align}

Then, we have $\epsilon_{CutMix}(\alpha)$ for DP-CutMixSL mechanism as follows:
\begin{align}
\epsilon_{CutMix}(\alpha) &= \frac{\alpha{(\max_{i\in\mathbf{C}}\lambda_i)}}{2}{\left(\frac{\Delta^2 D_s}{\sigma^2_s}+\frac{{(\max_{i\in\mathbf{C}}\lambda_i)}D_y}{\sigma^2_y}\right)}.
\label{eq:16}
\end{align}


\section{Robustness to Reconstruction Attack}

\begin{table}[h]
\caption{Privacy leakage measured by the reconstruction loss (MSE).}
\centering
    \resizebox{0.8\columnwidth}{!}{
    \begin{tabular}{l|rr}
    \toprule
    Type &   Train Dataset (10\%) & Train Dataset (100\%) \\ \cmidrule(l){1-3}
    Smashed data & 0.0091 & 0.0056\\ 
    Cutout& \textbf{0.0920} & \textbf{0.0829}\\ 
    Mixup & 0.0402 & 0.0351\\ 
    Patch CutMix & 0.0458 & 0.0434\\ 
    \bottomrule
    \end{tabular}}
\label{table:models}
\end{table}

Table 2 shows loss between raw data and reconstructed data generated from different types of mixing methods or datasets. To restore raw data from reconstructed data, we utilize a decoder model, comprised of two convolutional layers with additional interpolation methods to adaptively match the dimension to the aimed data size. For comparison, we train the decoder model with training datasets of two different sizes.
 
As a result, regardless of the training dataset size, the reconstruction loss was large in the following order: Cutout, patch CutMix, Mixup, and smashed data. In other words, it has robustness against reconstruction attacks in that order. Except for smashed data, Mixup is most vulnerable to reconstruction attack, because information leakage occurs over the entire area even though linear interpolation is taken. The proposed patch CutMix has relatively robustness since it can inject large-size noise into the local information necessary for reconstruction, thanks to its inherent masking, but is upper bounded on the Cutout, which discards a part of the image corresponding to the mask. Finally, the larger the training dataset size, the better the decoder model is trained, which reduces the overall reconstruction loss.


\end{document}